\documentclass{ws-procs961x669}            
\begin{document}
\title{Black Hole: The Interior Spacetime}

\author{Yen Chin Ong}

\address{Nordita, KTH Royal Institute of Technology \& Stockholm University,\\
Roslagstullsbacken 23, SE-106 91 Stockholm, Sweden\\
E-mail: yenchin.ong@nordita.org}

\begin{abstract}
The information loss paradox is often discussed from the perspective of the observers who stay outside of a black hole. However, the interior spacetime of a black hole can be rather nontrivial. We discuss the open problems regarding the volume of a black hole, and whether it plays any role in information storage. We also emphasize the importance of resolving the black hole singularity, if one were to resolve the information loss paradox. 
\end{abstract}


\bodymatter

\section{What's Inside a Black Hole?}

The information loss paradox remains unresolved 40 years after Hawking first pointed out the problem \cite{Hawking:1976ra}. Consider a cloud of matter in a pure state that collapses to form a black hole. The common viewpoint is that, if physics is indeed unitary, exterior observers should be able to recover a pure state after the black hole has completely evaporated. Indeed, according to black hole complementarity \cite{Susskind:1993if}, an exterior observer and an in-falling observer can observe different events, but each see a completely self-consistent, unitary-preserving, physics. It may therefore be sufficient to discuss the information loss paradox solely from the point of view of the exterior observers. Nevertheless, the fate of the information, as seen from a comoving observer that falls into the black hole, remains an interesting problem. To answer this question, one has to first ask the obvious question: ``\emph{what is inside a black hole?}'' We shall discuss some possibilities.

For simplicity, let us focus our attention on the Schwarzschild black hole, described by the following metric, in the units $G=c=1$,
\begin{equation}
g[\text{Sch}]=-\left(1-\frac{2M}{r}\right)\text{d}t^2 + \left(1-\frac{2M}{r}\right)^{-1}\text{d}r^2  +r^2(\text{d}\theta^2 +\sin^2\theta~ \text{d}\phi).
\end{equation}

A textbook on general relativity typically mentions that one can analytically continue the Schwarzschild manifold to the Kruskal-Szekeres manifold, which contains another asymptotically flat region inside the black hole, on the other side of the Einstein-Rosen bridge. There are at least two issues with this picture. The first is well known: in a realistic gravitational collapse, one does not seriously expect that the resulting black hole would contain another universe inside it. The second issue is that, in general, analytic continuations are not unique. If we drop some conditions such as vacuum, then other analytic continuations exist \cite{jose}. (Analyticity is a rather strong condition, if one drops this and considers smooth continuations only, then even more extensions are possible.)  
This means that the interior geometries of two black holes can be vastly different even if both of them are exactly Schwarzschild as seen from the outside.

Let us first consider the simplest scenario in which the black hole is formed from gravitational collapse and is thus ``one-sided'', i.e., does not harbor another asymptotically flat region. Can we say something definitive about its interior geometry? In particular, what is its volume? 

This is a rather tricky question to answer. Although the area of the black hole event horizon, which by virtue of being a null surface, is a well-defined geometric invariant independent of the choice of spacelike hypersurfaces, the same is not true for the spatial 3-volume inside the black hole\cite{brandon}. In fact, various definitions of black hole volumes have been proposed (see e.g. Ref.\refcite{1310.1935}), including definition that is more thermodynamical than geometrical \cite{1012.2888,1301.5926}. It is also noted that an invariant way of defining volume does exist, albeit this is not a proper volume \cite{0508108}. What we are interested in is the question: if we look at, say, the black hole at the center of the Milky Way, how much volume does it contain?

Since the proper volume is hypersurface-dependent, there is no unique answer to this. What one can ask, however, is the \emph{largest} volume possible. Assuming the Schwarzschild geometry, Christodoulou and Rovelli showed that \cite{1411.2854} a
cross section of the event horizon taken
at late times bounds a spatial volume which grows with time as
\begin{equation}
\text{Vol.} \sim 3\sqrt{3}\pi M^2 v,
\end{equation}
where $v$ is the advanced time, and $M$ the mass of the black hole. 
This volume corresponds to the maximal slice\cite{bruce, frank} at $r=3M/2$, which we shall refer to as the ``CR-volume''.
More explicitly, this is obtained via the volume integral
\begin{equation}\label{int1}
\text{Vol.} \sim \int^v \int_{S^2}  \max\left[r^2 \sqrt{\frac{2M}{r}-1}\right]  \sin\theta ~\text{d}\theta~ \text{d}\phi ~\text{d}v.
\end{equation}
The time dependence is not so surprising since the interior geometry of a Schwarzschild black hole is, unlike the exterior geometry, \emph{not} static. 

Christodoulou and Rovelli estimated that Sagittarius $\text{A}^*$, the supermassive black hole at the center of our galaxy, 
contains sufficient space to fit a \emph{million} solar systems, despite its
areal radius being only a factor of 10 or so larger than the distance from Earth to the Moon. Taking into account the rotation of Sagittarius $\text{A}^*$ and repeating the calculation using the Kerr metric, Bengtsson and Jakobsson showed that this estimate does not change by much \cite{1502.01907}, and so there is a lot of room inside a black hole!
(Such volume for asymptotically locally anti-de Sitter black hole has also been investigated \cite{Ong:2015tua}.)
It is therefore tempting to ask whether this large volume can shed some light on the information loss paradox.

\section{Black Hole Volume and Information Storage}

To answer this question, one must first ask: How much do we trust the naive spherically symmetric Schwarzschild geometry to continue to hold in the interior spacetime? 

Near the singularity it is conjectured that spacetime should become highly chaotic (BKL singularity \cite{BKL}), and one would not be able to trust the aforementioned naive volume integral to continue to hold. However, this complication should not arise until the very late part of the evolution. Therefore, let us for the moment assume that the Schwarzschild geometry is a sufficiently good approximation and there is indeed a large volume inside a black hole, at least for some time during its evolution under Hawking evaporation. The question is then: how does Hawking evaporation affect this volume? Does it shrink together with the horizon area? 

For the most part of the evolution, the mass loss is well-approximated by the thermal loss equation, which is given by 
\begin{equation}\label{ODE1}
\frac{\text{d}M}{\text{d}v} \approx -\alpha a\sigma T^4 = - \alpha a \left(27\pi M^2\right) \left(\frac{\hbar}{8\pi M}\right)^4 =  -\frac{\mathcal{C} }{M^2},  ~~\mathcal{C} > 0,
\end{equation}
where $a=\pi^2/(15 \hbar^3)$ is the radiation constant (which is $4/c$ times the Stefan-Boltzmann constant), and $\alpha$ is the greybody factor, which depends on the number of particle species emitted by the Hawking radiation. It is found in Ref.\refcite{1503.08245} that somewhat counter-intuitively, the CR-volume continues to increase even though the black hole is losing mass, and consequently, its area is decreasing. This is easy to see, for the volume integral is 
\begin{equation}
V(v) \sim 3\sqrt{3}\pi \int^v M^2 ~\text{d}v.
\end{equation}
(We remark that the volume is no longer asymptotically linear in $v$, as it would be when the mass of the black hole is a constant.)
So by the Fundamental Theorem of Calculus, one immediately sees that 
\begin{equation}
\frac{\text{d}V}{\text{d}v} \sim  3\sqrt{3} \pi M^2(v) \geqslant 0.  
\end{equation}
As discussed in Ref.\refcite{1503.08245}, this could imply that at sufficiently late time, one either have a ``bag-of-gold'' type of geometry, which is still connected to the exterior asymptotically flat spacetime via a throat, or that the throat could pinch off entirely and the interior spacetime becomes an isolated universe on its own. This is consistent with the results of Ref.\refcite{9405007}.

In view of this, it is perhaps tempting to conjecture that information can be stored in the CR-volume despite the shrinking of the black hole area. However, this appears not to be the case since the actual entropy content associated with the CR-volume, $S_{\text{CR}}$, turns out to be proportional to the \emph{area} (though the coefficient is not the same as the Bekenstein-Hawking area formula), instead of the volume. Explicitly, Zhang\cite{1510.02182} showed that 
\begin{equation}
S_{\text{CR}} \sim \frac{3\sqrt{3}}{\mathcal{C}(90 \pi \times 8^4)}A.
\end{equation}
Zhang also argued that the thermodynamics associated with the entropy
in the CR volume is caused by the vacuum polarization near the horizon. 

\section{The Importance of Understanding the Singularities}

One crucial issue that we have not discussed thus far is the singularity inside a black hole. For a Schwarzschild black hole this is particularly important, since it is spacelike and lies in the future of anything that falls into the hole. It is well known that the maximum proper time before an in-falling object eventually terminates at the singularity is $\pi M$. This means that if the singularity is not resolved, information would eventually ``fall off the edge of spacetime'' at the singularity. More precisely, there is an entanglement between particles behind the horizon, and particles that remain in the exterior of it. However, the one behind the horizon ultimately gets destroyed at the singularity -- hence the information loss paradox.

This means that having a large spacelike volume \emph{by itself} does not resolve the information loss paradox. Instead, one has to understand whether the singularity is indeed resolved in quantum gravity, and if so how. Although the common viewpoint is that the singularity is a sign that general relativity breaks down and will be cured by a full working theory of quantum gravity, \emph{it may not be}. 


Only if there is no singularity, then a black hole remnant with a huge interior volume or a baby universe may be a viable candidate to resolve the information loss paradox. This proposal is of course not without problems. The readers should refer to Ref.\refcite{Chen:2014jwq} for detailed discussion, but let us mention one of the obvious problem: If information is indeed contained in the large volume in some way or another (i.e. not necessarily in the CR-volume), this seems to violate the common wisdom that the Bekenstein-Hawking entropy -- which is proportional to the area -- should count the number of states of the black hole degrees of freedom. Nevertheless, there remains a possibility that the common wisdom is incorrect. For more discussion, see Refs.\refcite{Chen:2014jwq}, \refcite{0810.4886}, \refcite{0706.3239}, \refcite{0908.1265}, \refcite{1304.3803}.

\section{Conclusion: Pay Attention to Geometry}

To conclude, general relativity is a \emph{geometric} theory of gravity. We should therefore pay more attention to the spacetime geometry, whether it is the exterior geometry or the interior one. Let us not be biased by the fact that we are exterior observers. Perhaps the interior spacetime has nontrivial geometry that allows information to be stored despite the shrinkage of the black hole horizon during Hawking evaporation.

In addition, the volume of (two-sided) black holes has recently gained some attention in the context of holography. To be more specific, there might exist a volume/complexity relation, such that the computational complexity\cite{1402.5674} $C(t)$ of a certain quantum state, as a function of some proper time $t$, goes like
\begin{equation}
C(t)  \propto \text{Vol}(\Sigma_t),
\end{equation}
where $\Sigma_t$ is a codimension-one space-like section of the anti-de Sitter bulk with extremal volume. See Ref.\refcite{1510.00349} and the references therein for detail.

Despite the potentially important roles of black hole volumes, we should not ignore the singularities, since if they remain unresolved by quantum gravity information can be lost by simply getting destroyed there. 

In the context of the firewall controversy \cite{Almheiri:2012rt}, it has been proposed by Susskind that perhaps the firewall, if it exists, is just the singularity of the black hole that has ``migrated'' to the horizon, due to the volume inside a black hole gradually disappearing as the entanglement between the interior and exterior spacetime gets broken \cite{1208.3445} at sufficiently late times. This further exemplifies the importance of understanding both the volume, as well as the singularity, of a black hole. 

It is of course possible that the information loss paradox is a question that can be resolved without knowing the details whether -- or how -- quantum gravity cures the black hole singularity, but again, \emph{it may not be}.


\bibliographystyle{ws-procs961x669}

\end{document}